\documentclass[12pt]{article}
\usepackage{graphicx}
\def \b{{\cal B}}

\def \beq{\begin{equation}}
\def \beqn{\begin{eqnarray}}

\def \eeq{\end{equation}}
\def \eeqn{\end{eqnarray}}

\def \ite{{\it et al.}}
\def \ob{\overline{B}^0}

\def \ok{\overline{K}^0}

\def \v#1#2{V_{#1#2}}

\def\lsim{\mathrel{\rlap{\lower3pt\hbox{$\sim$}}\raise2pt\hbox{$<$}}}
\def\gsim{\mathrel{\rlap{\lower3pt\hbox{$\sim$}}\raise2pt\hbox{$>$}}}

\begin{document}
\renewcommand{\thetable}{\Roman{table}}
\rightline{ANL-HEP-PR-03-005-Rev}
\rightline{EFI-03-04-Rev}
\rightline{hep-ph/0302094}
\rightline{March 2003}
\bigskip
\bigskip
\centerline{\bf NEW PHYSICS CONTRIBUTIONS TO THE $B \to \phi K_S$ DECAY
  \footnote{Submitted to Phys.~Rev.~D.}}
\bigskip
\centerline{\it Cheng-Wei Chiang}
\centerline{\it High Energy Physics Division, Argonne National Laboratory,
Argonne, IL 60440}
\centerline{and}
\centerline{\it Enrico Fermi Institute, University of Chicago, Chicago, IL
  60637}
\medskip

\centerline{\it Jonathan L. Rosner}
\centerline{\it Enrico Fermi Institute and Department of Physics}
\centerline{\it University of Chicago, Chicago, IL 60637}
\bigskip
\centerline{\bf ABSTRACT}
\medskip
\begin{quote}
  Recent measurements of the time-dependent $CP$ asymmetry of the $B \to \phi
  K_S$ decay give results whose central values differ from standard model
  expectations.  It is shown how such data can be used to identify new physics
  contributions in a model-independent manner.  In general, a sizeable new
  amplitude with nontrivial weak and strong phases would be required to explain
  current data.  Improvement in the quality of data will allow one to form a
  more definite conclusion.
\end{quote}
\medskip
%\leftline{\qquad PACS codes: 13.25.Hw, 11.30.Hv, 14.40.Nd, 13.75.Lb}
\newpage

\centerline{\bf I.  INTRODUCTION}
\bigskip

It has been suggested to look for discrepancies among the time-dependent $CP$
asymmetries of different $B$ decay modes as a means to detect new physics
\cite{Grossman:1996ke,Fleischer:1996bv,London:1997zk,Fleischer:2001pc,%
Hiller:2002ci,Ciuchini:2002pd}.  Since the $B \to J/\psi K_S$ decay is a
tree-dominated process in the standard model (SM), its $CP$ asymmetry ${\cal
S}_{J/\psi K_S}$ is believed to be less affected by new physics and to give
information on $\sin 2\beta$.  Although the $CP$ asymmetry of the $B \to \phi
K_S$ mode is also expected to give the same $\sin 2\beta$ within the SM, this
process is, however, particularly sensitive to new physics contributions
because it is a purely penguin loop-mediated process in the SM.  The SM
pollution from a small $u$-penguin with the weak phase $\gamma$ has been
studied in Ref.~\cite{Grossman:1997gr} and it is found that the deviation of
${\cal S}_{\phi K_S}$ from $\sin 2\beta$ is of ${\cal O}(\bar \lambda^2) \sim
5\%$, where $\bar\lambda \simeq {\cal O}(0.2)$ is a parameter close in
magnitude to the Wolfenstein parameter $\lambda \simeq 0.22$
\cite{Wolfenstein:1983yz}.  Therefore, a large deviation of ${\cal S}_{\phi
  K_S}$ from its SM prediction would signal contributions from physics beyond
the SM.

As argued by Fleischer and Mannel \cite{Fleischer:2001pc}, even if one ignores
rescattering effects, contributions from new physics at a TeV scale to $\Delta
I=0$ operators could be of the same order as the SM ones, while new $\Delta
I=1$ operators are suppressed by $\bar\lambda$.  With rescattering effects
taken into account, both the new $\Delta I=1$ operators and the SM pollution
will be enhanced by about $\bar\lambda$.  In any case, both of the $\Delta
I=0,1$ operators from the new TeV-scale physics can be more significant in
comparison with the above-mentioned SM pollution.

The world average of $\sin 2\beta$ as measured from the golden mode $B \to J/
\psi K_S$, $\sin 2\beta = 0.734 \pm 0.054$ \cite{NirICHEP02}, agrees well with
constraints obtained from other experiments.  Recently both the BaBar and Belle
groups have also reported measurements of time-dependent $CP$ asymmetries in
the $B \to \phi K_S$ decay.  ${\cal S}_{\phi K_S}$ (the coefficient of $\sin
\Delta m t$ in flavor-tagged decays) is found to be about $2.7 \sigma$ away
from ${\cal S}_{J/\psi K_S}$, while ${\cal A}_{\phi K_S}$ (the coefficient of
$\cos \Delta m t$) is $1 \sigma$ away from $0$.  If this situation continues as
the data precision improves, it would be of interest to know the magnitude and
phase of possible new physics contributions to the $\phi K_S$ mode.

Instead of separating the new physics contributions into $\Delta I=0$ and
$\Delta I=1$ parts as done in Ref.~\cite{Fleischer:2001pc}, we will simplify
the discussion by considering the combined amplitude from such effects along
with the smaller SM pollution amplitude.  This enables us to obtain useful
information from the three observables ${\cal S}_{\phi K_S}$, ${\cal A}_{\phi
  K_S}$, and the ratio $R$ between the sum of squared amplitudes extracted from
the measured $B^0 (\ob) \to \phi K^0 (\ok)$ branching ratios and a ``standard''
squared amplitude, such as the SM predicted value or experimentally measured
$B^{\pm} \to K^* \pi^{\pm}$ branching ratio.  The algebraic structure of the
problem then becomes very similar to that studied by several authors \cite{GR}
for $B \to \pi \pi$.  We try to find in a model-independent way the allowed
magnitude and phases of the new amplitude and some generic properties
associated with it.  Such an analysis is useful in helping us narrow down new
physics models \cite{Datta:2002nr} consistent with observed data.

The paper is organized as follows.  Sec.\ II introduces a decomposition of
decay amplitudes in terms of topological contributions.  The formalism for
time-dependent CP asymmetries is discussed in Sec.\ III.  We present numerical
analyses for two separate cases of new physics in Secs.\ IV and V.  In Sec.\
VI, we summarize our results.

\bigskip
\centerline{\bf II.  TOPOLOGICAL AMPLITUDE ANALYSIS}
\bigskip

In the framework of the SM, both the $\phi K^0$ and $\phi K^{\pm}$ modes
receive important contributions from QCD and EW penguin graphs, with the former
having a dominant effect.  Useful information about the QCD penguin
contribution can be obtained from the $K^{*0} \pi^{\pm}$ decay mode using
flavor-$SU(3)$ symmetry \cite{Chiang:2001ir}.  It should be noted that a tiny
annihilation diagram also exists in both the $\phi K^{\pm}$ and $K^{*0}
\pi^{\pm}$ decay modes.  From the arguments of both dynamical suppression and
the fact that no asymmetry is observed between the $K^{*0} \pi^{\pm}$ modes, we
shall ignore the annihilation amplitude in these charged decays.  In this case,
both the neutral and charged $\phi K$ modes have the same decay amplitudes.  We
will then average over the branching ratios of these two sets of modes using
their associated errors as the weights for our analysis.

Let's write down the amplitudes of the relevant modes in terms of independent
topological components as \cite{Chau:pn,Gronau:1994rj}
\beqn
{\cal A}(\phi K^0)
&=& p \, e^{i (\phi_{SM} + \delta_p)} + s \, e^{i (\phi_{SM} + \delta_s)} ~,
\label{eq:phiK0} \\
{\cal A}(K^{*0} \pi^+)
&=& p \, e^{i (\phi_{SM} + \delta_p)} ~,
\label{eq:Kstar0pi}
\eeqn
in the SM.  In the above two equations, the $p$ part denotes the QCD penguin
contribution which also contains a negligible color-suppressed EW penguin
amplitude, and the $s$ part denotes the EW penguin contribution along with a
small flavor-$SU(3)$-singlet amplitude, as expected from the OZI rule.  The
variables $p$ and $s$ are absolute values of the respective amplitudes and
therefore are non-negative by definition.  The weak phase $\phi_{SM}$
satisfying $e^{-2i\phi_{SM}} = V_{tb} V_{ts}^* / (V_{tb}^* V_{ts})$ is the same
for both the $p$ and $s$ parts.  Finally, $\delta_p$ and $\delta_s$ are the
associated strong phases.  Note that to simplify the notation given in
Ref.~\cite{Chiang:2001ir}, we omit from these amplitudes the subscript $P$
indicating that the spectator quark ends up in the pseudoscalar meson in the
final state and the prime that denotes $\Delta S = 1$ transitions.

It should be noted that we explicitly assume flavor $SU(3)$ symmetry in
Eqs.~(\ref{eq:phiK0}) and (\ref{eq:Kstar0pi}) in order to relate the amplitude
for the $K^{*0} \pi^+$ mode to the penguin part in the $\phi K$ process in
later analysis.  The $SU(3)_F$ breaking effect will be characterized by the
factor $[f_{\phi} F^{B \to K}(m_{\phi}^2)] / [f_{K^*} F^{B \to \pi}(m_{K^*}^2)]
\sim 1.2$.  We will simply treat this extra factor as $1$ in our analysis.

New physics can give rise to new operators that contribute to the decays of the
above processes.  We will distinguish two cases in later discussions: (i) only
the $\phi K$ modes receive the new contributions while the $K^{*0} \pi^{\pm}$
modes are purely SM processes; and (ii) both types of decay modes receive the
same contributions from new physics.  Case (i) could happen, for example, when
new physics enters the $b \to s$ EW penguin only.  In this case, we add an
extra amplitude $n \, e^{i (\phi_n + \delta_n)}$ to Eq.~(\ref{eq:phiK0}).  Case
(ii) could happen when new physics modifies the $b \to s$ QCD penguin.  In that
case we add the new amplitude to both Eq.~(\ref{eq:phiK0}) and
Eq.~(\ref{eq:Kstar0pi}).  In general, the new amplitude $n \, e^{i (\phi_n +
  \delta_n)}$ is a combination of $\Delta I = 0$ and $\Delta I = 1$ ones that
may contribute at different strengths \cite{Fleischer:2001pc}.  A more careful
job can in principle be done by separating the new amplitude into those with
different isospins and studying new physics contributions in each piece.
However, one would find that there are not enough observables among the decay
modes to solve for all the parameters in the amplitudes.

\bigskip
\centerline{\bf III.  TIME-DEPENDENT $CP$ ASYMMETRIES}
\bigskip

In this section, we review the general analysis of time-dependent $CP$
asymmetry of pure $B^0$ and $\ob$ decays into a $CP$ eigenstate $f_{CP}$.
Let's define the asymmetry as
\beq
a_{f_{CP}}(t) \equiv
\frac{\Gamma\left(\ob_{\rm phys}(t) \to f_{CP}\right) 
      - \Gamma\left(B^0_{\rm phys}(t) \to f_{CP}\right)}
     {\Gamma\left(\ob_{\rm phys}(t) \to f_{CP}\right) 
      + \Gamma\left(B^0_{\rm phys}(t) \to f_{CP}\right)} ~.
\eeq
In our case, $f_{CP} = \phi K_S$.  Denote
\beq
\label{eq:lambda_phiKs}
\lambda_{\phi K_S} =
\eta_{\phi K_S} \left( \frac{q}{p} \right)_B \left( \frac{p}{q} \right)_K
\frac{{\overline {\cal A}}(\phi \ok)}{{\cal A}(\phi K^0)} ~,
\eeq
where $\eta_{\phi K_S} = -1$ is the $CP$ eigenvalue of the $\phi K_S$ state,
\beq
\left( \frac{q}{p} \right)_B = \frac{V_{tb}^* V_{td}}{V_{tb} V_{td}^*}
\quad \mbox{and} \quad
\left( \frac{p}{q} \right)_K = \frac{V_{cs} V_{cd}^*}{V_{cs}^* V_{cd}}
\eeq
are factors that account for the mixing effects in neutral $B$ and $K$ meson
systems, respectively, and
\beqn
\label{eq:A_Abar}
{\cal A}(\phi K^0)
&\equiv& \langle K^0 | {\cal H} | B^0 \rangle
= a \, e^{i (\phi_a + \delta_a)} + b \, e^{i (\phi_b + \delta_b)} ~, \\
{\overline {\cal A}}(\phi \ok)
&\equiv& \langle \ok | {\cal H} | \ob \rangle
= a \, e^{i (-\phi_a + \delta_a)} + b \, e^{i (-\phi_b + \delta_b)} ~,
\eeqn
where $a,b$ are chosen to be positive, $\phi_{a,b} \in \{ -\pi,\pi \}$ and
$\delta_{a,b} \in \{ 0,2\pi \}$ are the associated weak and strong phases,
respectively.  The above amplitudes are invariant under the transformations
$\phi_{a,b} \to \phi_{a,b} \pm m \pi, \delta_{a,b} \to \delta_{a,b} \mp m \pi$
and $\phi_{a,b} \to \phi_{a,b} \pm m \pi, \delta_{a,b} \to \delta_{a,b} \pm m
\pi$ for $m \in {\bf Z}$.  Here the separation of the total amplitude into two
parts is done in accord with the nature of the problem.  The ratio of the
amplitudes in Eqs.\ (6) and (7) is then
\beq
\frac{{\overline {\cal A}}(\phi \ok)}{{\cal A}(\phi K^0)}
= e^{-2i\phi_a}
  \frac{1 + r e^{i(\phi - \delta)}}{1 + r e^{-i(\phi + \delta)}} ~,
\eeq
where 
\beq
r \equiv b/a \ge 0~~,~~~ \phi \equiv \phi_a-\phi_b~~,~~~{\rm and}
~~~\delta \equiv \delta_a - \delta_b~~~.
\eeq
The $CP$ asymmetry can then be written as
\beq
a_{\phi K_S} (t)
= {\cal A}_{\phi K_S} \cos (\Delta M t)
  + {\cal S}_{\phi K_S} \sin (\Delta M t) ~,
\eeq
where $\Delta M$ is the mass difference between the two physical $B$ meson
states, and
\beqn
{\cal A}_{\phi K_S}
&=& \frac{|\lambda_{\phi K_S}|^2 - 1}{|\lambda_{\phi K_S}|^2 + 1} ~, \\
{\cal S}_{\phi K_S}
&=& \frac{2 \, {\rm Im} \lambda_{\phi K_S}}{|\lambda_{\phi K_S}|^2 + 1} ~.
\eeqn

New physics will affect the $CP$ asymmetry observables through the parameter
$\lambda_{\phi K_S}$.  Therefore, it may come in at two places: the mixing
matrix and/or the decay amplitudes.  As emphasized in
Ref.~\cite{Grossman:1996ke}, new physics effects on the mixing part will be
universal and do not change the SM predicted pattern of $CP$ asymmetries in
different modes; their effects on the decay amplitudes, however, are
non-universal so that the $CP$ asymmetries can vary from channel to channel.
Since current $\sin 2\beta$ measurements from other decay modes, such as
$J/\psi K_S$, $\eta' K_S$, etc, seem to agree with one another and with the
unitarity triangle constraints obtained from other processes pretty well, it is
plausible to assume that any strange behavior in the $\phi K_S$ mode is mostly
due to new physics contributions in the amplitudes.  In this case, we will use
the SM mixing factors in Eq.~(\ref{eq:lambda_phiKs}) and obtain
\beq
\label{eq:simp_lambda}
\lambda_{\phi K_S}
= - e^{-2 i \beta_{\rm eff}}
  \frac{1 + r e^{i(\phi - \delta)}}{1 + r e^{-i(\phi + \delta)}} ~,
\eeq
where
\beq
e^{-2 i \beta_{\rm eff}}
= \left( \frac{q}{p} \right)_B \left( \frac{p}{q} \right)_K
  e^{-2i\phi_a} ~.
\eeq
Within the SM, $\phi_a \simeq \pi$ and one obtains the effective weak phase
$\beta_{\rm eff}$ coinciding with $\beta$ in the unitarity triangle.  However,
if new physics modifies the phase $\phi_a$, then $\beta_{\rm eff}$ will in
general differ from what the SM expects.  If one writes $\phi_a = \phi_{SM} +
\phi$ with $\phi_{SM}$ being the phase expected in the SM and $\phi$ being the
deviation, then $\beta_{\rm eff} = \phi_{SM} + \phi$ (mod $\pi$).

We will be exclusively dealing with the decays of a $B$ meson into a final
state with one pesudoscalar meson ($P$) and one vector meson ($V$).  The
invariant amplitude ${\cal A}$ of such a process is conventionally related to
its partial width in the following way:
\beq
\Gamma(B \to PV) = \frac{(p^*)^3}{8 \pi m_B^2}|{\cal A}(B \to PV)|^2 ~,
\eeq
where $p^*$ is the $3$-momentum of each final particle in the rest frame of the
$B$ meson, and $m_B$ is the mass of the decaying $B$ meson.  Note that $p^*$ is
raised to its third power to appropriately account for the P-wave kinematic
factor.

\bigskip
\centerline{\bf IV.  NEW PHYSICS ONLY IN $\phi K$ SYSTEM}
\bigskip

In this section, we will discuss the case when new physics only enters the
$\phi K$ system but not the $K^* \pi$ system.  Since the $p$ and $s$ parts of
the $\phi K$ decay amplitudes have the same weak phase $\phi_{SM}$, we can
combine them into a single $v$ part and write, including the new physics part,
\beq
\label{eq:NPphiK}
{\cal A}(\phi K^0)
= v \, e^{i (\phi_{SM} + \delta_v)} + n \, e^{i (\phi_n + \delta_n)}
= v \, e^{i (\phi_{SM} + \delta_v)}
  \left[ 1 + r \, e^{-i (\phi + \delta)} \right] ~,
\eeq
where $r = n / v$, $\phi = \phi_{SM} - \phi_n$, and $\delta = \delta_v -
\delta_n$.  Since here we assume that the $K^{*0} \pi^{\pm}$ modes are not
affected by the new physics, we can use them to obtain reliable information on
$p$, the magnitude of the QCD penguin.  The effective Hamiltonian approach
indicates that there is a small relative strong phase between the $p$ and $s$
amplitudes in the SM \cite{Fleischer:1993gr,Ali:1998eb}.  The relative strong
phase obtained in this approach comes purely from short-distance physics
\cite{Bander:px}.  In general, there are nonperturbative strong phases from
soft gluon exchanges in the final-state particles that may be different between
the two types of penguin diagrams.  For simplicity and definiteness, we will
take $\delta_v = \delta_p \simeq \delta_s - \pi \simeq 0$ since the overall
strong phase will not matter, and consider maximal destructive interference
between the QCD and EW penguins, in accord with the effective Hamiltonian
analysis.  The additional $-\pi$ in the above strong phase relation does not
really have a strong interaction origin but simply comes from the charge
coupling of the final-state $s$ quark with the $Z$ boson in the EW penguin.
Therefore, the $s$ part has a $180^{\circ}$ phase from the $p$ part within the
SM.  Under this assumption, $v = p - s$ can be computed once we know the
prediction of the ratio $s / p$ in the SM.  We will also mention consequences
of imperfect destructive interference between these two types of amplitudes.

\bigskip
\leftline{\bf A.  Observables}
\bigskip

One observable in the $\phi K$ system is the ratio
\beq
\label{eq:R}
R
\equiv \frac{|{\cal A}^{\rm exp}(B^0 \to \phi K^0)|^2
               + |{\cal A}^{\rm exp}(\ob \to \phi \ok)|^2}
            {2 |{\cal A}^{\rm SM}(B^0 \to \phi K^0)|^2}
= 1 + 2 r \cos\phi \cos\delta + r^2 ~.
\eeq
The numerator in the above definition is the sum of measured branching ratios
of $B^0 \to \phi K^0$ and $\ob \to \phi \ok$.  As described before, we will
actually take the weighted average of the neutral and charged modes.  The
denominator is the theoretical prediction for the same branching ratio sum
within the SM.  In terms of $R$ and Eq.~(\ref{eq:simp_lambda}), we obtain
\beqn
\label{eq:RS}
R \, {\cal S}_{\phi K_S}
&=& \sin 2\beta + 2\,r \cos\delta\sin(2\beta - \phi) + r^2 \sin2(\beta - \phi)
~, \\
\label{eq:RA}
R \, {\cal A}_{\phi K_S}
&=& 2\,r \sin\phi \sin\delta ~.
\eeqn
Now we have three observables $R$, ${\cal S}_{\phi K_S}$, and ${\cal A}_{\phi
  K_S}$ that allow us to solve for the three parameters $r$, $\phi$, and
$\delta$.  As the value of $R$ may vary owing to the interference between $p$
and $s$, we will estimate the SM contribution and also search the allowed
parameter space by varying $R$ over a reasonable range.

The self-tagging modes $B^\pm \to \phi K^\pm$ can provide additional
statistical power to the determination of ${\cal A}_{\phi K_S}$ if we assume
that ${\cal A}(B^+ \to \phi K^+) = {\cal A}(B^0 \to \phi K^0)$ as we have done
above.  In that case one finds just
\beq
A_{CP} \equiv \frac{|{\cal A}(\phi K^-)|^2 - |{\cal A}(\phi K^+)|^2}
{|{\cal A}(\phi K^-)|^2 + |{\cal A}(\phi K^+)|^2} = {\cal A}_{\phi K_S}
= 2\,r \sin\phi \sin\delta/R
\eeq
for the time-integrated CP rate asymmetry.
The BaBar Collaboration \cite{Aubert:2003tk} has recently reported
$A_{CP} = 0.039 \pm 0.086 \pm 0.011$.  We shall not use this value in our
averages but in principle it can greatly reduce the error on ${\cal A}_{\phi
K_S}$.

\bigskip
\leftline{\bf B.  Numerical studies}
\bigskip

In this subsection, we will use the measured values of ${\cal A}_{\phi K_S}$,
${\cal S}_{\phi K_S}$, and $R$ with some theoretical input from the SM to find
the allowed ranges of $r$, $\phi$, and $\delta$.  Solving for $r$ in
Eq.~(\ref{eq:R}) in terms of $R$, $\phi$, and $\delta$, one obtains two
solutions
\beqn \label{eqn:rvals}
r_1
&=& - \cos\phi \cos\delta - \sqrt{\cos^2\phi \cos^2\delta + R - 1} ~,
\quad \mbox{(Solution I)}
\nonumber \\
r_2
&=& - \cos\phi \cos\delta + \sqrt{\cos^2\phi \cos^2\delta + R - 1} ~.
\quad \mbox{(Solution II)}
\eeqn
First, it is seen that Solution I is not allowed for $R > 1$ because $r_1$ has
to be positive.  Therefore, we see that $r_1 < 1$.

% This is Table I
\begin{table}
\caption{Experimental input of measured branching fractions.
\label{tab:BR}}
\begin{center}
\begin{tabular}{cccc} \hline \hline
($\times 10^{-6}$) & 
  ${\cal B}(\phi K^0)$ & ${\cal B}(\phi K^+)$ & ${\cal B}(K^{*0} \pi^+)$ \\
\hline
CLEO & 
  $5.4^{+3.7}_{-2.7}\pm0.7 \, (<12.3)$ \cite{Briere:2001ue} & 
  $5.5^{+2.1}_{-1.8}\pm0.6$ \cite{Briere:2001ue} & 
  $7.6^{+3.5}_{-3.0}\pm1.6$ \cite{Jessop:2000bv} \\
BaBar & 
  $7.6^{+1.3}_{-1.2}\pm0.5$ \cite{Aubert:2003tk} &
  $10.0^{+0.9}_{-0.8}\pm0.5$ \cite{Aubert:2003tk} &
  $15.5\pm3.4\pm1.8$ \cite{Aubert:2001ap} \\
Belle & 
  $10.0^{+1.9+0.9}_{-1.7-1.3}$ \cite{ChenICHEP02} & 
  $10.7\pm1.0^{+0.9}_{-1.6}$ \cite{ChenICHEP02} & 
  $19.4^{+4.2+2.1+3.5}_{-3.9-2.1-6.8}$ \cite{Abe:2002av} \\
\hline
Average &
  $7.98\pm1.07$ & $9.51\pm0.78$ & $12.3\pm2.5$ \\
\hline \hline
\end{tabular}
\end{center}
\end{table}

In the effective Hamiltonian approach, the EW penguin is found to be of
considerable importance, with the ratio $|s / p|$ predicted to be between
$10\%$ and $11\%$ using the results given in Ref.~\cite{Ali:1998eb}.  On the
other hand, the $B^+ \to K^{*0} \pi^+$ decay mode involves only $p$, ignoring a
small annihilation diagram that also contributes to the $\phi K^+$ mode.  Using
its branching ratio, we obtain $|p| = (1.42 \pm 0.14) \times 10^{-8}$.
Combining the above results and assuming maximal destructive interference
between $p$ and $s$, the SM predicts $|{\cal A}^{\rm SM}(\phi K)| = (1.27 \pm
0.13) \times 10^{-8}$.  To improve the statistics, we take the weighted average
for the branching ratios of the neutral and charged $\phi K$ modes as given in
Table \ref{tab:BR} and obtain $|{\cal A}^{\rm exp}(\phi K)| = (1.27 \pm 0.04)
\times 10^{-8}$ after removing the kinematic factors.  Therefore, we obtain an
estimate of
\beq
R = 
\left| \frac{{\cal A}^{\rm exp}(\phi K)}{{\cal A}^{\rm SM}(\phi K)} \right|^2 
\simeq 0.99 \pm 0.21 ~.
\eeq
If a nontrivial relative strong phase exists between $p$ and $s$, the central
value of the resulting $R$ will become smaller.  In the case of maximal
constructive interference between $p$ and $s$, $R$ could be as low as $0.5$.

% This is Table II
\begin{table}
\caption{Experimental input of measured $CP$ asymmetries for the $B \to \phi
  K_S$ mode.
\label{tab:CPA}}
\begin{center}
\begin{tabular}{cccc} \hline \hline
Quantity & BaBar \cite{HameldeMonchenault} & Belle \cite{Abe:2002np} & Average \\
\hline
${\cal S}$ & $-0.18\pm0.51\pm0.07$ & $-0.73\pm0.64\pm0.22$ & $-0.38\pm0.41$ \\
${\cal A}$ & $0.80\pm0.38\pm0.12$ & $-0.56\pm0.41\pm0.16$ & $0.19\pm0.30$ \\
\hline \hline
\end{tabular}
\end{center}
\end{table}

We will use the $CP$ asymmetry observables measured by the BaBar and Belle
groups \cite{HameldeMonchenault,Abe:2002np} as given in Table \ref{tab:CPA} for
our analysis.  Replacing $r$ in Eqs.~(\ref{eq:RS}) and (\ref{eq:RA}) by one of
the above solutions, it is then possible to find on the $\phi$-$\delta$ plane
regions that are consistent with the measured values ${\cal A}_{\phi K_S} =
0.19 \pm 0.30$, ${\cal S}_{\phi K_S} = -0.38 \pm 0.41$
\cite{HameldeMonchenault,Abe:2002np}, and the additional requirement that $r
\ge 0$ by definition.  The fact that ${\cal A}_{\phi K_S}$ is negative at the
$1 \sigma$ level gives the following possibilities: (i) $-\pi \le \phi < 0$ and
$0 \le \delta < \pi$; and (ii) $0 \le \phi < \pi$ and $\pi \le \delta < 2\pi$.

% This is Figure 1
\begin{figure}
\begin{center}
\includegraphics[height=2.4in]{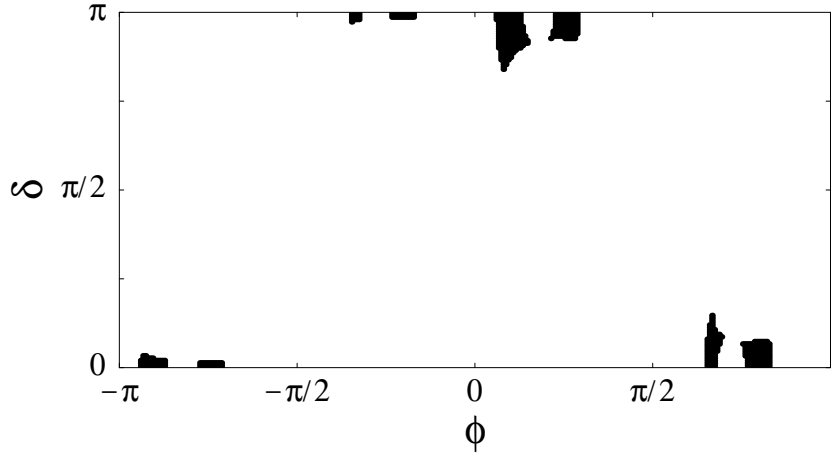} \\
(a) \\
\vspace{0.2cm}
\includegraphics[height=2.4in]{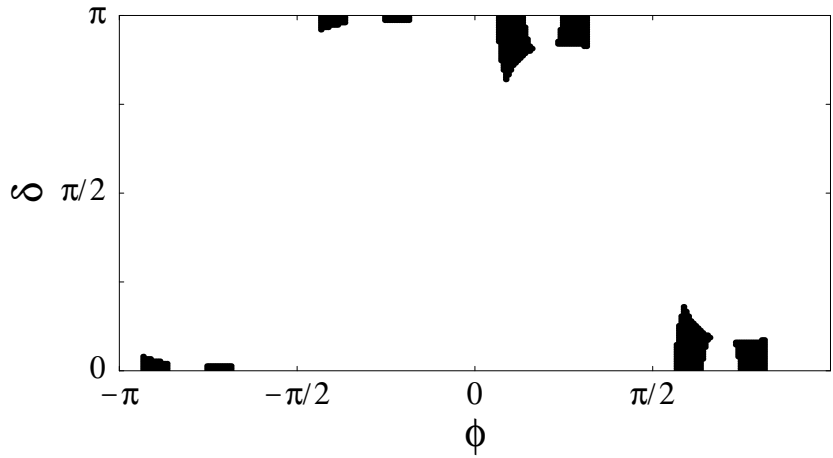} \\
(b) \\
\vspace{0.2cm}
\includegraphics[height=2.4in]{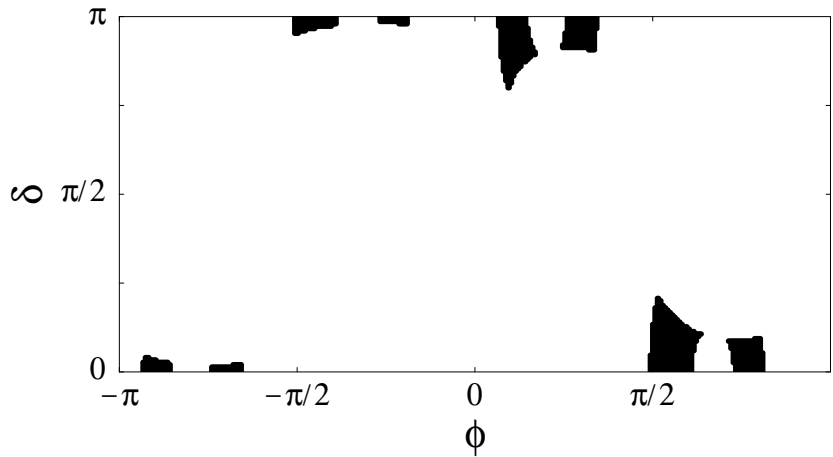} \\
(c)
\caption{The allowed regions in the $\phi$-$\delta$ plane for Solution II with
  (a) $R = 0.8$, (b) $R = 1.0$, and (c) $R = 1.2$.  Here we only show the
  allowed regions in the range $-\pi \le \phi \le \pi$, $0 \le \delta \le \pi$.
  Other regions can be obtained by $\phi \to \phi \pm \pi$ and $\delta \to
  \delta \pm \pi$.  Solution I is not allowed for $R \gsim 0.8$ and, therefore,
  no corresponding plots are shown here.  For $R=1$ ${\cal S}_{\phi K_S}$ and
  ${\cal A}_{\phi K_S}$ are unchanged under $\delta \to \pi - \delta$ and $\phi
  \to \phi + \frac {\pi}{2}$ [Fig.\ 1(b)].
\label{fig:phase}}
\end{center}
\end{figure}

In Fig.~\ref{fig:phase}, we only show a set of representative solutions in the
range $-\pi \le \phi \le 0$, $0 \le \delta \le \pi$ for $R = 0.8, 1.0$ and
$1.2$.  It is noticed that Solution I does not exist when $R \gsim 0.8$.
Therefore, we only show those for Solution II.  As shown in Section III,
solutions in other regions on the $\phi$-$\delta$ plane can be obtained by the
translations $\phi \to \phi \pm \pi$ and $\delta \to \delta \pm \pi$.  As $R$
increases, the allowed regions of Solution II become larger.

% This is Figure 2
\begin{figure}
\begin{center}
\includegraphics[height=2.7in]{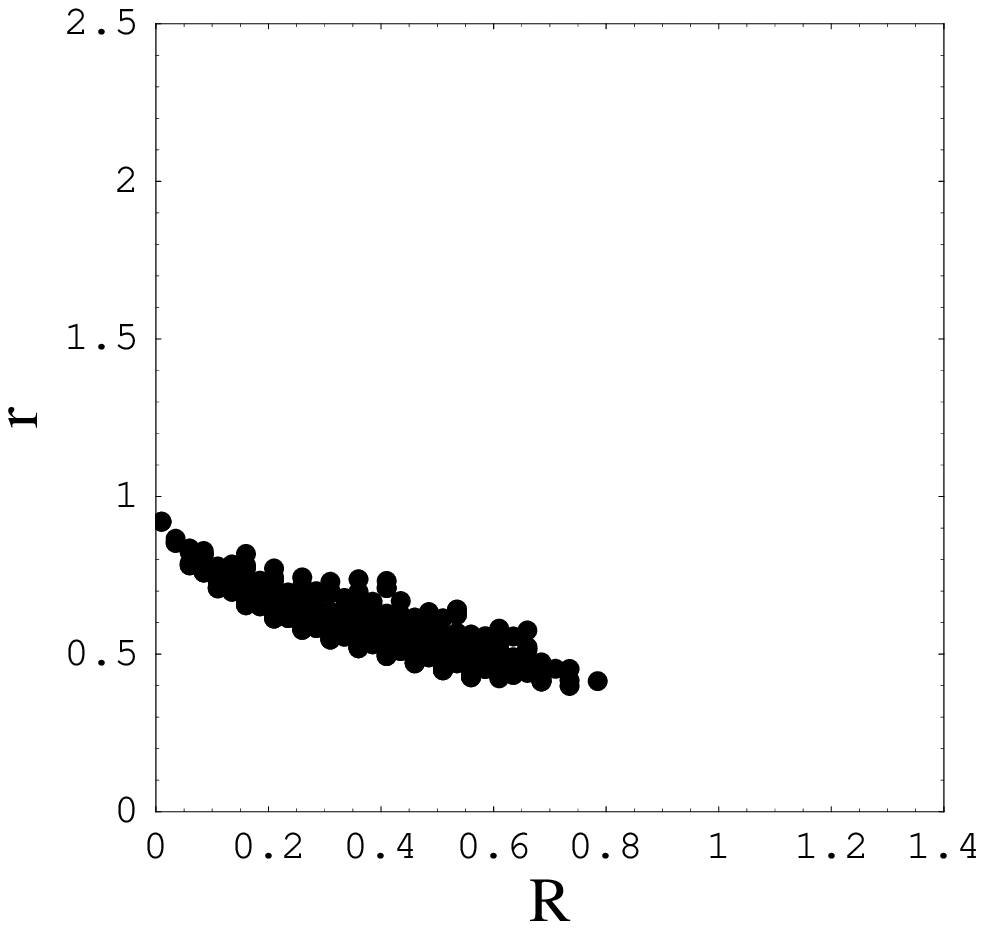}\hspace{0.5cm}
\includegraphics[height=2.7in]{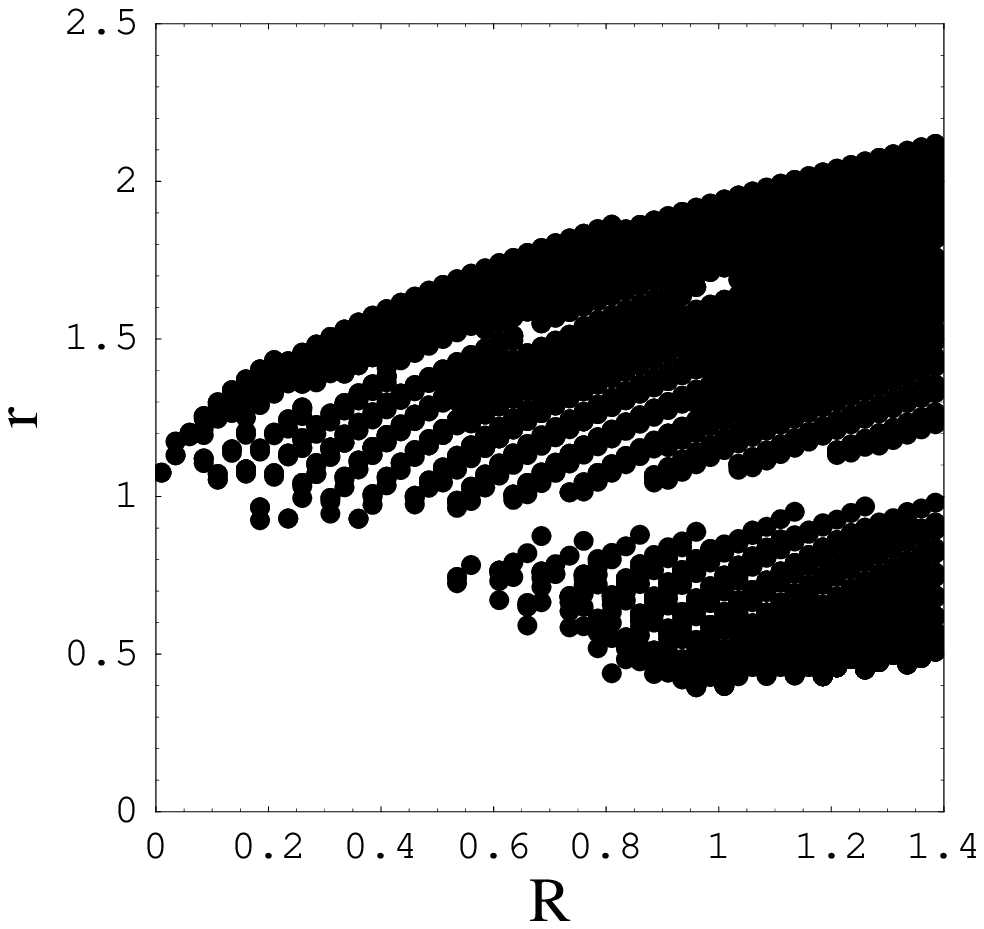} \\
(a) \hspace{6.8cm} (b)
\caption{The allowed range of $r$ for specific values of $R$, using (a)
  Solution I and (b) Solution II.
\label{fig:r}}
\end{center}
\end{figure}

In Fig.~\ref{fig:r}, we plot the allowed ranges of $r$ for $0 \le R \le 1.4$.
The dark region in plot (a) corresponds to Solution I, and that in plot (b) to
Solution II.  It is seen from the plots that to satisfy the constraints of
measured data, $r$ has to be at least about $0.4$ for either solution.  This
corresponds to a new physics amplitude with a magnitude of at least about $0.45
\times 10^{-8}$.  It is also found that for Solution I, $R$ has to be less than
about 0.8.  Therefore, the current value of $R$ favors Solution II.  If we take
the value $R = 1$, Solution II has a wide range for $r$: 
$0.40 \lsim r \lsim 0.90$ and $1.05 \lsim r \lsim 1.96$.

The fact that $r$ has to be greater than a minimum can be readily understood.
Should $r$ be too small, then new physics [the $n$ part in
Eq.~(\ref{eq:NPphiK})] does not have enough weight to change $R {\cal S}_{\phi
  K_S}$ from that extracted from the $J/\psi K_S$ mode to the measured one as
the modification is of ${\cal O}(r)$ according to Eq.~(\ref{eq:RS}).  As
mentioned in the beginning, $r \sim {\cal O}(1)$ means that the new amplitude
has the same order of size as the SM contribution.  This would point to the
possibility of new physics at the TeV scale or below.

\begin{figure}
\begin{center}
\includegraphics[height=5.8in]{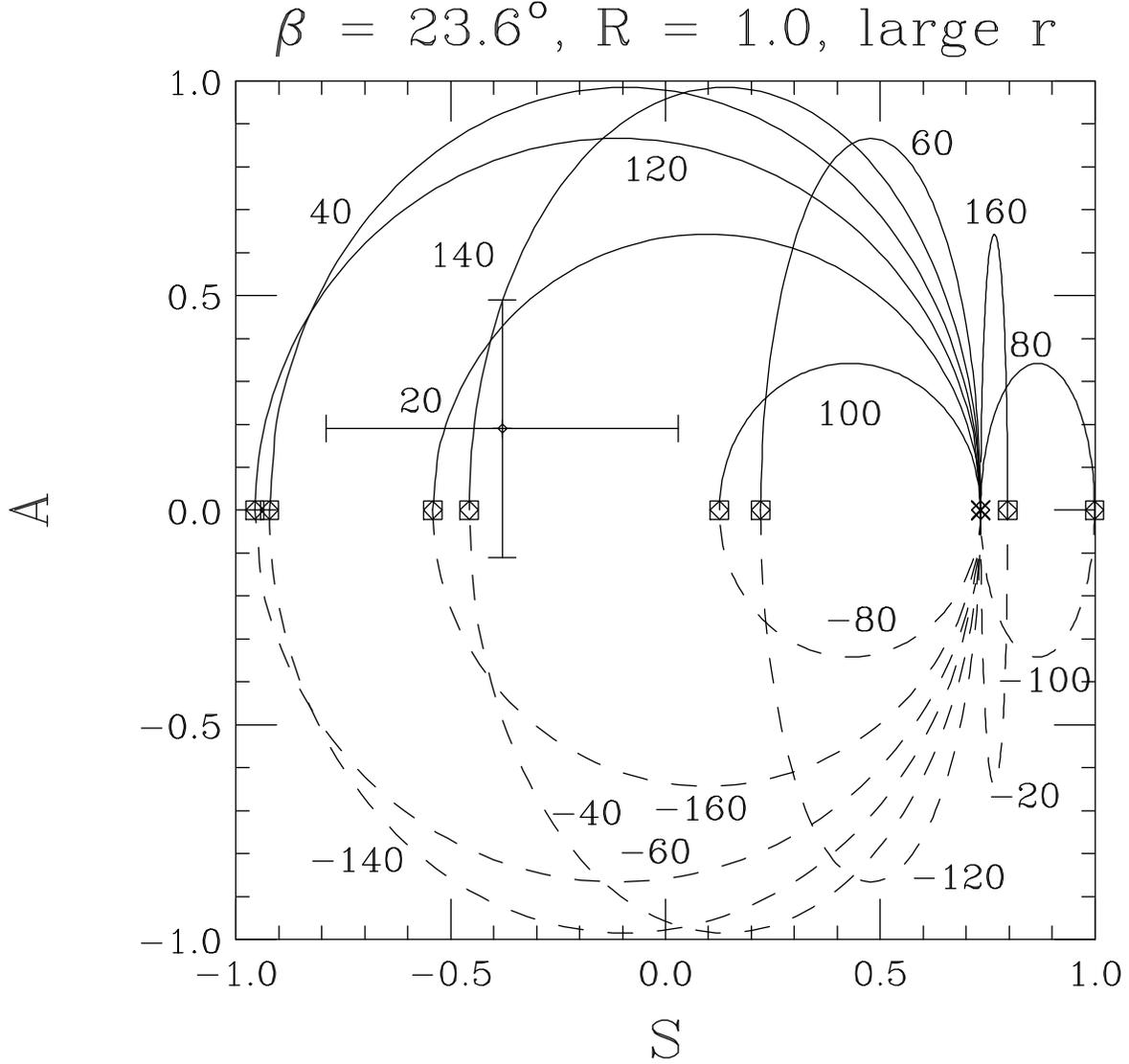}
\caption{The curves traced in the ${\cal S}_{\phi K_S}$-${\cal A}_{\phi K_S}$
  plane by varying the relative strong phase $\delta$ between 0 and $\pi$ for
  fixed values of $\phi$.  The plot is for $\beta = 23.6^{\circ}$ and $R=1$
  with $r$ chosen according to Solution II; no solution I exists for $R=1$.
  Curves are labeled by values of $\phi$ (dashed: $\phi < 0$; solid: $\phi >
  0$) in degrees.  Squares and diamonds correspond to values of $\delta = 0$ or
  $\pi$.  The point at ${\cal S}_{\phi K_S} = 0.734$, ${\cal A}_{\phi K_S} = 0$
  corresponds to $\phi = 0,~\pm \pi$ for all $\delta$.  The plotted data point
  is the average quoted in Table II.
\label{fig:SA}}
\end{center}
\end{figure}

The sensitivity of ${\cal S}_{\phi K_S}$ and ${\cal A}_{\phi K_S}$ to the weak
and strong phases $\phi$ and $\delta$ for a value of $R$ close to the central
one is illustrated in Fig.\ \ref{fig:SA}.  Here each curve for a given $\phi$
intersects the axis ${\cal A}_{\phi K_S} = 0$ at either $\delta = 0$ or $\delta
= \pi$, while curves with ${\cal A}_{\phi K_S} < 0$ are related to those with
${\cal A}_{\phi K_S} > 0$ and the same value of ${\cal S}_{\phi K_S}$ by the
transformation $\phi \to \pi + \phi$, $\delta \to \pi - \delta$.  The plotted
cross shows the present status of the data summarized in Table II.  The ranges
of $\delta$ and $\phi$ are restricted in general by the requirement that the
argument $\cos^2 \phi \cos^2 \delta + R - 1$ of the square roots in Eqs.\ 
(\ref{eqn:rvals}) be non-negative.  Because of the special value of $R$, we are
able to draw curves for essentially any value of $\phi$ by varying $\delta$.
Therefore, the constraints come merely from ${\cal S}_{\phi K_S}$ and ${\cal
  A}_{\phi K_S}$.

% This is Figure 3

\begin{figure}
\begin{center}
\includegraphics[height=2.9in]{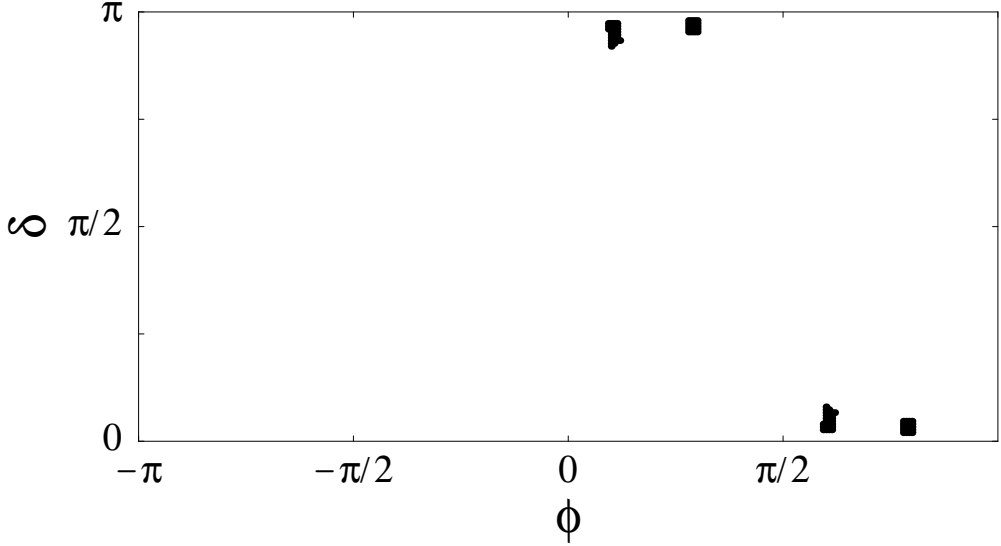}
\caption{The allowed regions on the $\phi$-$\delta$ plane for $R = 1.0$ using
  Solutions II and with a factor of 3 improvement in ${\cal A}_{\phi K_S}$ and
  ${\cal S}_{\phi K_S}$.
\label{fig:smallphase}}
\end{center}
\end{figure}

Assuming the central values of ${\cal S}_{\phi K_S}$ and ${\cal A}_{\phi K_S}$
stay the same in future experiments but the errors are improved by a factor of
3, we find that the allowed regions become smaller.
This is demonstrated in Fig.~\ref{fig:smallphase}.

% This is Figure 4

We find that the variation of $\sin 2\beta$ within its experimental range makes
little difference in the allowed solutions.  The general behavior and regions
presented in Figs.~\ref{fig:phase} and \ref{fig:r} remain the same.

\bigskip
\centerline{\bf V.  NEW PHYSICS IN $\phi K$ and $K^* \pi$ SYSTEMS}
\bigskip

In this section, we consider the situation of new physics entering both the
$\phi K$ and $K^* \pi$ systems.  Since the new physics contribution $n \, e^{i
  (\phi_n + \delta_n)}$ is to be added to both Eqs.~(\ref{eq:phiK0}) and
(\ref{eq:Kstar0pi}), we combine the $p$ and $n$ parts into a single $q$ part as
follows
\beqn
{\cal A}(\phi K^0)
&=& q \, e^{i (\phi_q + \delta_q)} + s \, e^{i (\phi_{SM} + \delta_s)}
= q \, e^{i (\phi_q + \delta_q)} 
    \left[ 1 + r' \, e^{-i (\phi + \delta)} \right] ~,
\label{eq:phiK2} \\
{\cal A}(K^{*0} \pi^+)
&=& q \, e^{i (\phi_q + \delta_q)} ~,
\label{eq:Kstar0pi2}
\eeqn
where $q \, e^{i (\phi_{q} + \delta_q)} = p \, e^{i (\phi_{SM} + \delta_p)} + n
\, e^{i (\phi_n + \delta_n)}$, $r' = s / q$, $\phi = \phi_q - \phi_{SM}$, and
$\delta = \delta_q - \delta_s$.  As mentioned in the previous section,
$\phi_{SM} \simeq \pi$ and $\delta_s \simeq \pi$, one thus should use $(\phi_q,
\delta_q) \simeq (\phi + \pi, \delta + \pi)$ to obtain the weak and strong
phases associated with the $q$ part when interpreting our following plots drawn
on the $\phi$-$\delta$ plane.

\bigskip
\leftline{\bf A.  Observables}
\bigskip

In this case, we use the observable
\beqn
R'
&\equiv& \frac{|{\cal A}^{\rm exp}(B^0 \to \phi K^0)|^2
                + |{\cal A}^{\rm exp}(\ob \to \phi \ok)|^2}
              {|{\cal A}^{\rm exp}(B^+ \to K^{*0} \pi^+)|^2
                + |{\cal A}^{\rm exp}(B^- \to {\overline K}^{*0} \pi^-)|^2}
\nonumber \\
&=& 1 + 2 r' \cos\phi \cos\delta + r'\,^2 ~,
\label{eq:Rprime}
\eeqn
Note that in spite of the similarity in the forms between $R'$ and $R$ defined
in the previous section, they are actually very different.  Using
Eq.~(\ref{eq:Rprime}), we have
\beqn
\label{eq:RprimeS}
R' \, {\cal S}_{\phi K_S}
&=& \sin 2\beta_{\rm eff} + 2\,r' \cos\delta \sin(2\beta_{\rm eff} - \phi)
    + r'\,^2 \sin 2(\beta_{\rm eff} - \phi) \nonumber \\
&=& \sin 2(\beta + \phi) + 2\,r' \cos\delta \sin(2\beta + \phi)
    + r'\,^2 \sin 2\beta ~, \\
\label{eq:RprimeA}
R' \, {\cal A}_{\phi K_S}
&=& 2\,r' \sin\phi \sin\delta ~,
\eeqn
where $\beta_{\rm eff} = \beta + \phi$ is used.  Here one quickly realizes that
we also have only three parameters, $r'$, $\phi$, and $\delta$ for which to
solve.

As in the previous case, the self-tagging rate asymmetry for $B^\pm \to \phi
K^\pm$ provides additional statistical power for the measurement of ${\cal
  A}_{\phi K_S}$, since
\beq
A_{CP} = 2\,r' \sin\phi \sin\delta/ R'~.
\eeq

\bigskip
\leftline{\bf B.  Numerical studies}
\bigskip

Solving $r'$ in Eq.~(\ref{eq:Rprime}) in terms of $R'$, $\phi$, and $\delta$,
one obtains two solutions
\beqn
r'_1
&=& - \cos\phi \cos\delta - \sqrt{\cos^2\phi \cos^2\delta + R' - 1} ~,
\quad \mbox{(Solution I)}
\nonumber \\
\label{eq:rprime}
r'_2
&=& - \cos\phi \cos\delta + \sqrt{\cos^2\phi \cos^2\delta + R' - 1} ~.
\quad \mbox{(Solution II)}
\eeqn
First, as in the previous case, Solution I is not allowed for $R' > 1$ because
$r'_1$ has to be positive.  Therefore, we see that $r'_1 < 1$.

% This is Figure 5
\begin{figure}
\begin{center}
\includegraphics[height=1.52in]{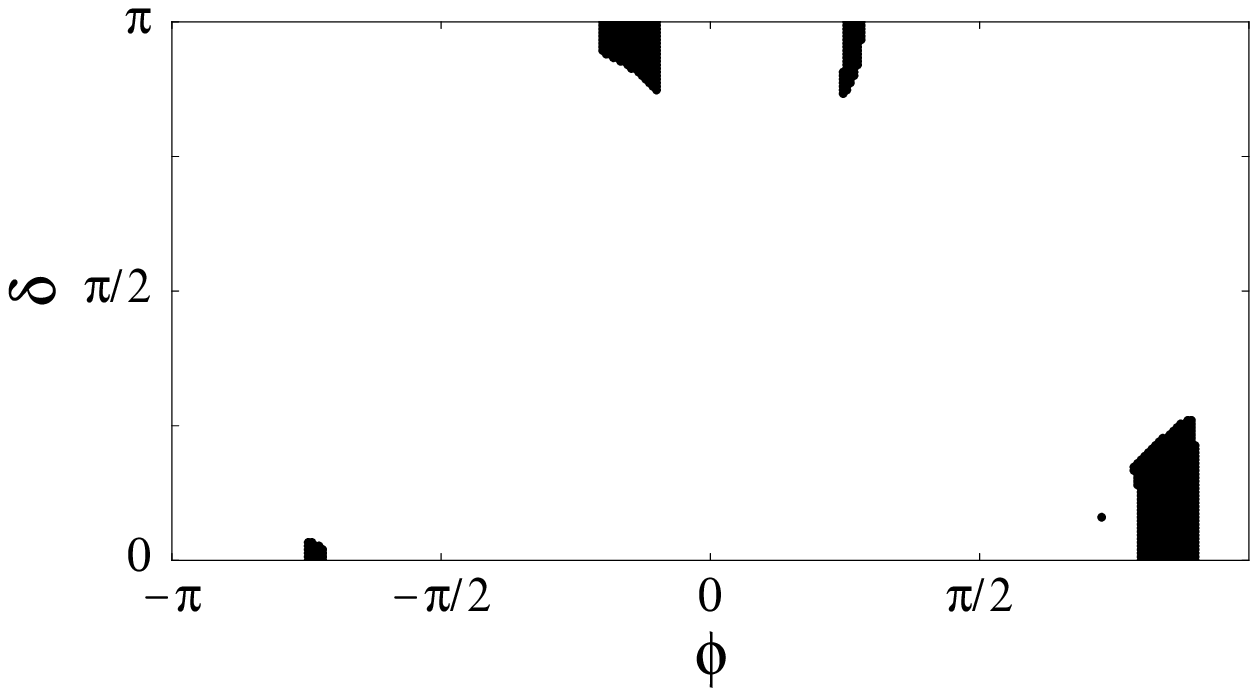}\hspace{0.1cm}
\includegraphics[height=1.52in]{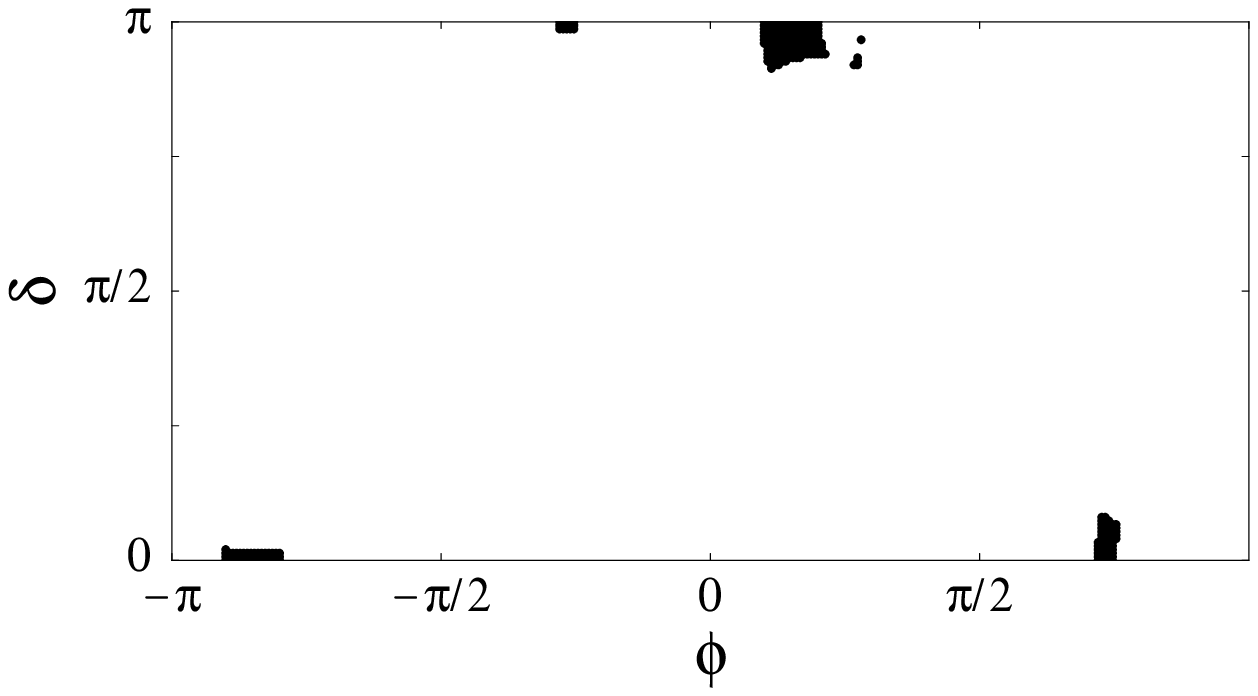} \\
(a) \hspace{7.5cm} (c) \\
\vspace{0.5cm}
\includegraphics[height=1.52in]{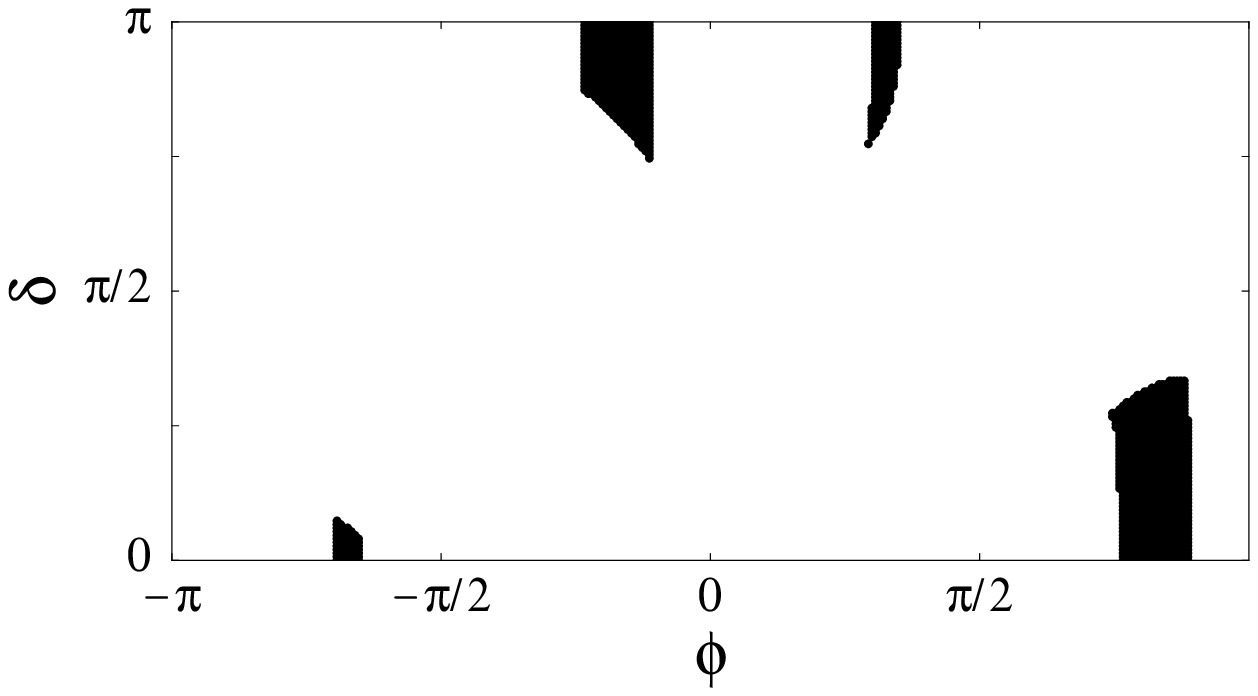}\hspace{0.1cm}
\includegraphics[height=1.52in]{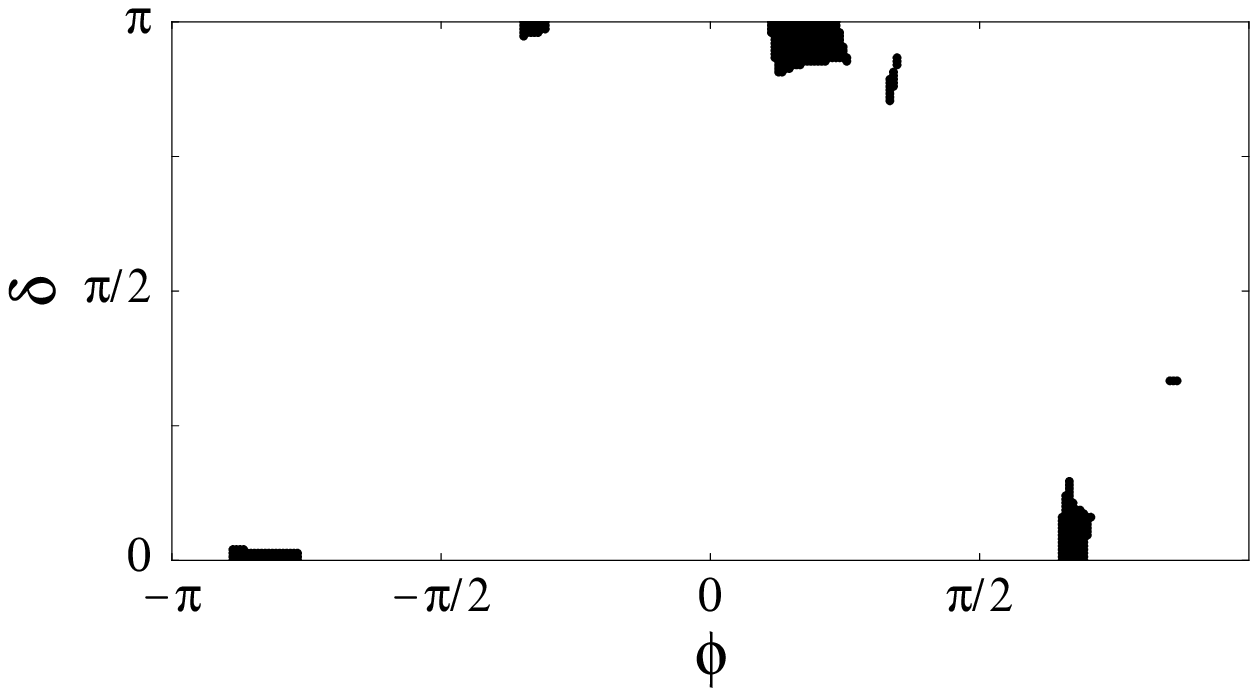} \\
(b) \hspace{7.5cm} (d) \\
\vspace{0.5cm}
\hspace{7.6cm}
\includegraphics[height=1.52in]{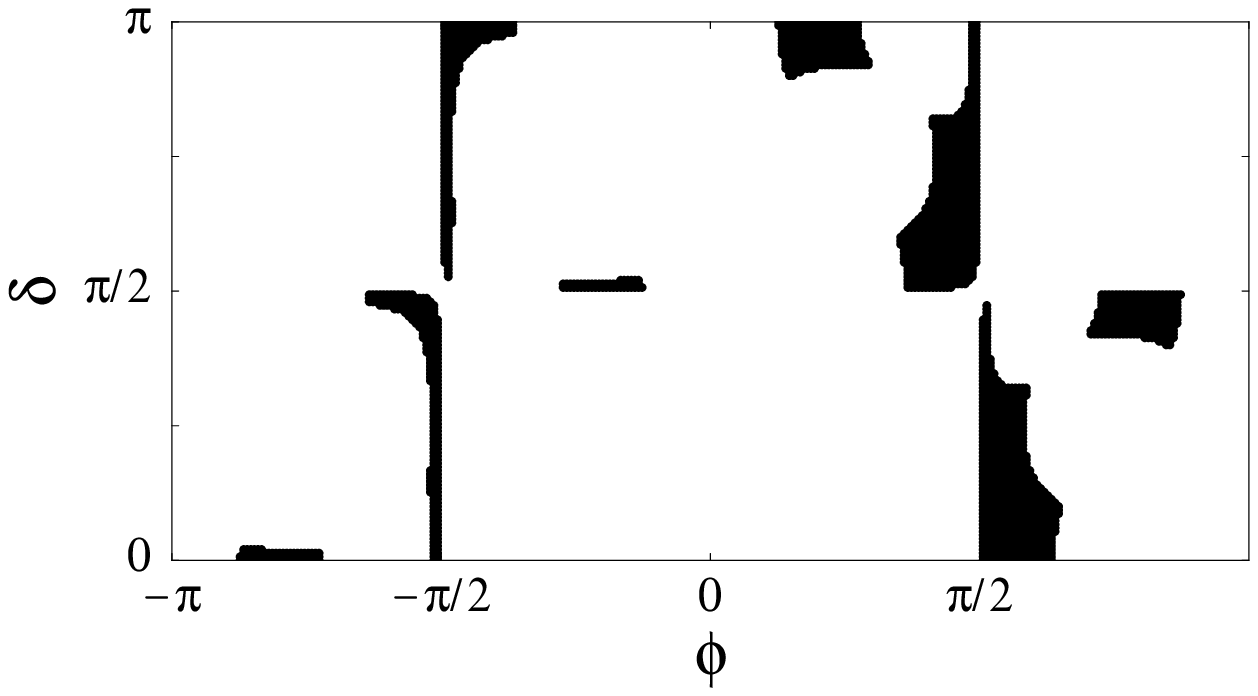} \\
\hspace{7.6cm} (e) \\
\caption{The allowed regions on the $\phi$-$\delta$ plane for $R' = 0.6, 0.8$
  and $1.0$, using Solutions I (plots (a) and (b) in the left column) and II
  (plots (c), (d), and (e) in the right column).  Here we only show the allowed
  regions in the range $-\pi \le \phi \le 0$, $0 \le \delta \le \pi$.  The
  other regions can be obtained by shifting: $\phi \to \phi + \pi$ and $\delta
  \to \delta + \pi$.  Solution I is not allowed for $R' = 1$.  In this case
  Fig.\ 5(e) displays an additional symmetry of the solution under $\delta \to
  \delta + \frac{\pi}{2}$, $\phi \to \pm \pi - \phi$.
\label{fig:phase2}}
\end{center}
\end{figure}

We extract the amplitude for the $K^* \pi$ mode from Table \ref{tab:BR} to be
$|{\cal A}^{\rm exp}(K^* \pi)| = (1.42 \pm 0.14) \times 10^{-8}$.  Combined
with the weighted average of the $\phi K$ mode amplitude size given in the
previous section, $|{\cal A}^{\rm exp}(\phi K)| = (1.25 \pm 0.05) \times
10^{-8}$, we find $R' = 0.79 \pm 0.17$.  In Fig.~\ref{fig:phase2}, we show a
set of representative solutions in the range $-\pi \le \phi \le 0$, $0 \le
\delta \le \pi$.  We take $R' = 0.6, 0.8$ and $1.0$.  It is seen that both
solutions have two allowed regions except for Solution I at $R' = 1.0$.  As
$R'$ increases, the allowed regions become larger for Solution II.

% This is Figure 6
\begin{figure}
\begin{center}
\includegraphics[height=2.7in]{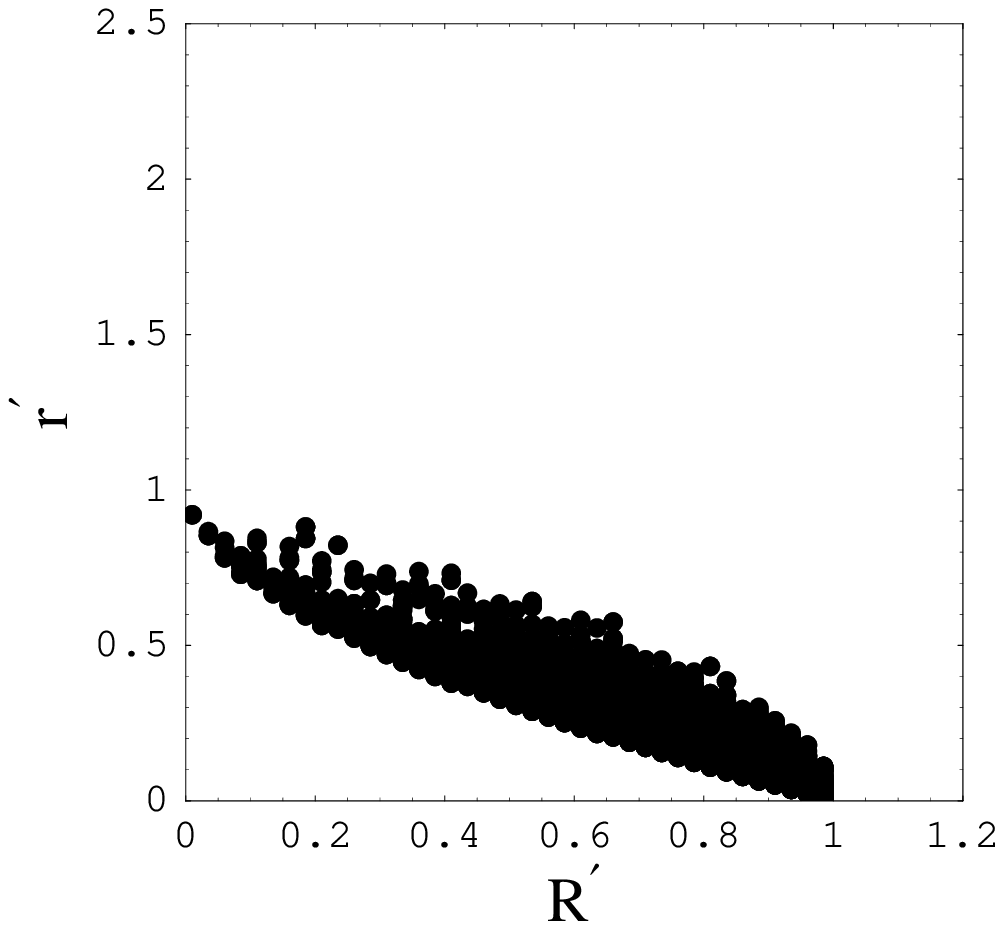}\hspace{0.5cm}
\includegraphics[height=2.7in]{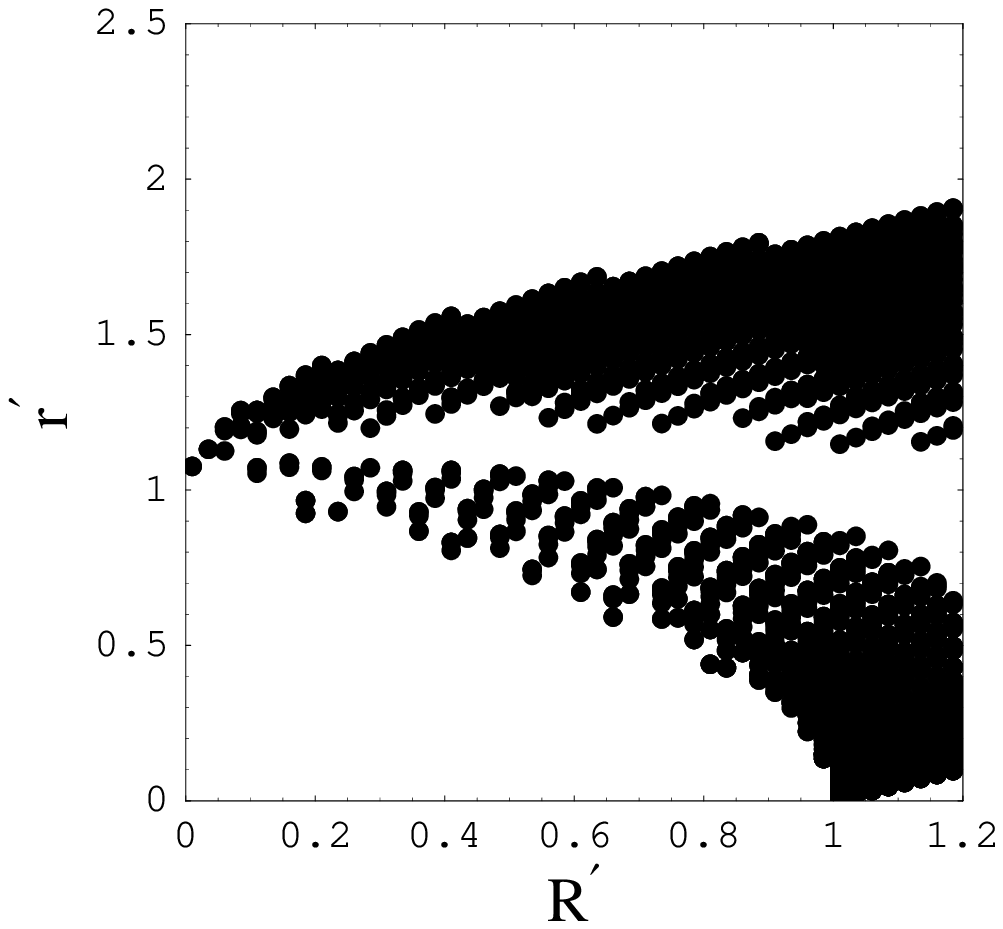} \\
(a) \hspace{6.8cm} (b)
\caption{The allowed range of $r'$ for specific values of $R'$, using (a)
Solution I and (b) Solution II.
\label{fig:r2}}
\end{center}
\end{figure}

In Fig.~\ref{fig:r2}, we also draw the allowed ranges of $r'$ for $0 \le R' \le
1.2$.  The dark region in plot (a) corresponds to Solution I, and that in plot
(b) to Solution II.  It is seen from the plots that to satisfy the constraints
of measured data, $r'$ can go down to almost $0$ for $R' \simeq 1$.  If we take
the value $R' = 0.8$, Solution I has $0.16 \lsim r' \lsim 0.44$ while Solution
II has a wider range $0.47 \lsim r' \lsim 0.97$ and $1.20 \lsim r' \lsim 1.76$.
In the standard model, one expects $r' \simeq 0.1$, $\phi = \pm \pi$, $\delta
\simeq \pm\pi$ in accord with the expected contribution (mentioned previously)
of the electroweak penguin amplitude.

If it turns out that $r' \ll 1$, that means the $n$ part and the $p$ part
interfere to give an amplitude larger in size, agreeing with the fact that $R'$
will be about $1$.  However, if $r' \sim {\cal O}(1)$, then there is a
cancellation between $n$ and $p$ such that the combined amplitude of the two
becomes comparable to the $SU(3)_F$-singlet amplitude.  Either situation would
tell us the new physics contribution is important.

\begin{figure}
\begin{center}
\includegraphics[height=2.8in]{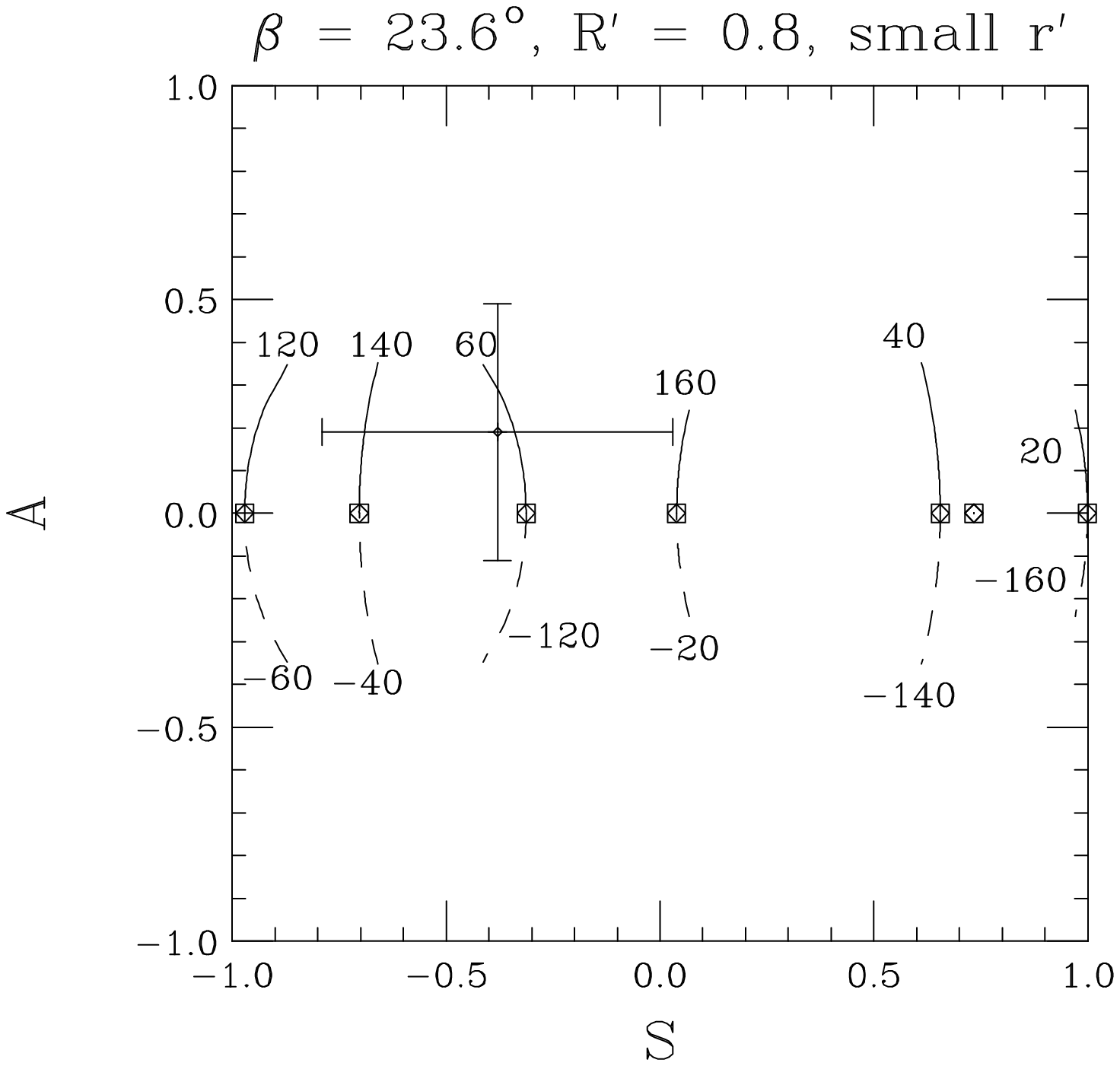}
\hspace{0.1cm}
\includegraphics[height=2.8in]{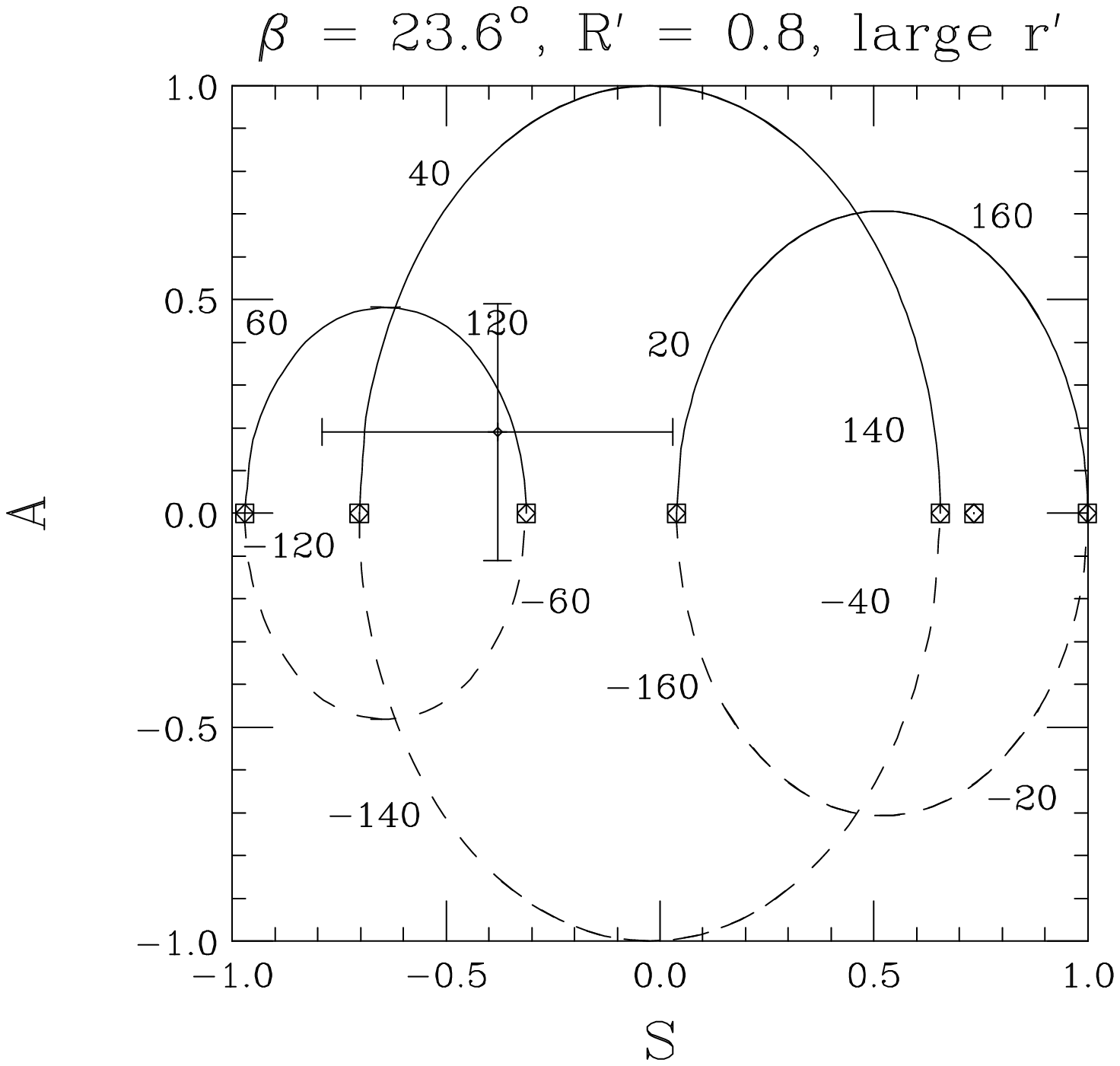} \\
(a) \hspace{7.3cm} (b)
\caption{Same as Fig.\ 3 but for new physics entering through the penguin
  penguin amplitude into both $B \to \phi K$ and $B^+ \to K^{*0} \pi^+$.  The
  left plot (a) is for Solution I and the right plot (b) for Solution II.  In
  both plots, $\beta = 23.6^{\circ}$ and $R'=0.8$.
\label{fig:SAprime}}
\end{center}
\end{figure}

The sensitivity of ${\cal S}_{\phi K_S}$ and ${\cal A}_{\phi K_S}$ to the weak
and strong phases $\phi$ and $\delta$ in Solutions I and II for the case in
which new physics enters through the penguin amplitude into both $B \to \phi K$
and $B^+ \to K^{*0} \pi^+$ is illustrated in Fig.\ \ref{fig:SAprime}.  As in
Fig.\ \ref{fig:SA}, each curve for a given $\phi$ intersects the axis ${\cal
  A}_{\phi K_S} = 0$ at either $\delta = 0$ or $\delta = \pi$, while curves
with ${\cal A}_{\phi K_S} < 0$ are related to those with ${\cal A}_{\phi K_S} >
0$ and the same value of ${\cal S}_{\phi K_S}$ by $\phi \to \pi + \phi$,
$\delta \to \pi - \delta$.  Contrary to Fig.\ \ref{fig:SA}, not all values of
$\phi$ and $\delta$ are allowed for making a curve with a given value of $R' <
1$.

It is interesting to notice that for the large-$r'$ solution (Solution II), the
curves for values of $\phi$ and $\pm \pi - \phi$ overlap, leading to the
appearance of continuity.  This is because both curves are part of a common
ellipse, obtained by solving Eqs.~(\ref{eq:Rprime}), (\ref{eq:RprimeS}), and
(\ref{eq:RprimeA}):
\beqn
\label{eq:ellipse}
\left( \frac{R' {\cal S}_{\phi K_S} - (\cos 2\phi + R' - 1) \sin 2\beta}
            {r_x} \right)^2
+ \left( \frac{R' {\cal A}_{\phi K_S}}{r_y} \right)^2
= \cos^2\phi + R' - 1 ~,
\eeqn
where $r_x = 2 \sin\phi \cos 2\beta$, and $r_y = 2 \sin\phi$.  These ellipses
have their centers at coordinates $([\cos 2\phi + R' - 1] \sin 2\beta / R',0)$.
One immediately sees that the above elliptic equation is invariant under the
transformation $\phi \to \pi - \phi$.  Since no explicit choice of solutions of
$r$ in Eq.~(\ref{eq:rprime}) is made for deriving Eq.~(\ref{eq:ellipse}), it is
valid for either solution.  This is why each curve associated with $\phi$ in
Fig.~\ref{fig:SAprime}(a) is actually a portion of the corresponding curve
associated with $\pi - \phi$ in Fig.~\ref{fig:SAprime}(b).  The curves for both
solutions are truncated (although not seen in the plot for Solution II because
of the overlap) because of the conditions $\cos^2\phi \cos^2\delta + R' - 1 \ge
0$, $r'_1 \ge 0$.

% This is Figure 7

Again, we find that if the experimental precision of ${\cal S}_{\phi K_S}$ and
${\cal A}_{\phi K_S}$ can be improved by a factor of $3$ with their current
central values, then the allowed regions are considerably restricted.  The
variation of $\sin 2\beta$ within its experimental range also does not affect
the general behavior and regions presented in this section.

\bigskip
\centerline{\bf VI.  SUMMARY}
\bigskip

We have shown how to estimate the magnitude and weak and strong phases of any
new physics contribution which might account for the deviation of the CP
asymmetry parameters in $B \to \phi K^0$ from their standard-model values.  We
find that it is useful to compare the overall rate for this process either with
that predicted in the standard model (a ratio $R$) or with that for $B^+ \to
K^{*0} \pi^+$ (a ratio $R'$), which is expected to be dominated by the penguin
amplitude in the standard model.

It is observed from our analysis that an amplitude with considerable size,
nontrivial weak and/or strong phases from new physics is required to fit the
current experimental results.  For example, in the case that new physics
contributes to the $\phi K$ modes but not the $K^* \pi$ mode, the ratio of the
new amplitude to the SM contribution has to be $\gsim 0.4$, independent of the
value of $R$.

Current experimental data indicate that the size $r$ or $r'$ of a new physics
amplitude relative to that of the standard model could well be of ${\cal O}(1)$
for a wide range of $R$ or $R'$.  Such ${\cal O}(1)$ parameters could indicate
new physics at about the TeV scale.  Of course, these extra contributions would
need the right strong and weak phases in order to explain the current data.
Considerable refinement of the rate and asymmetry measurements in $B \to \phi
K^0$ is necessary before the amplitude can be pinpointed satisfactorily,
however.

\bigskip
\centerline{\bf ACKNOWLEDGMENTS}
\bigskip

C.-W.~C. would like to thank R.~Briere and H.~C.~Huang for experimental
information.  This work was supported in part by the United States Department
of Energy, High Energy Physics Division, under Contract Nos. DE-FG02-90ER-40560
and W-31-109-ENG-38.
 
\bigskip

% Journal and other miscellaneous abbreviations for references
% Phys. Rev. D style
\def \ajp#1#2#3{Am.~J.~Phys.~{\bf#1}, #2 (#3)}
\def \apny#1#2#3{Ann.~Phys.~(N.Y.) {\bf#1}, #2 (#3)}
\def \app#1#2#3{Acta Phys.~Polonica {\bf#1}, #2 (#3)}
\def \arnps#1#2#3{Ann.~Rev.~Nucl.~Part.~Sci.~{\bf#1}, #2 (#3)}
\def \art{and references therein}
\def \b97{{\it Beauty '97}, Proceedings of the Fifth International
Workshop on $B$-Physics at Hadron Machines, Los Angeles, October 13--17,
1997, edited by P. Schlein}
\def \carg{{\it Masses of Fundamental Particles -- Carg\`ese 1996}, edited by
M. L\'evy \ite, NATO ASI Series B:  Physics Vol.~363 (Plenum, New York, 1997)}
\def \cmp#1#2#3{Commun.~Math.~Phys.~{\bf#1}, #2 (#3)}
\def \cmts#1#2#3{Comments on Nucl.~Part.~Phys.~{\bf#1}, #2 (#3)}
\def \corn93{{\it Lepton and Photon Interactions:  XVI International
Symposium, Ithaca, NY August 1993}, AIP Conference Proceedings No.~302,
ed.~by P. Drell and D. Rubin (AIP, New York, 1994)}
\def \cp89{{\it CP Violation,} edited by C. Jarlskog (World Scientific,
Singapore, 1989)}
\def \dpff{{\it The Fermilab Meeting -- DPF 92} (7th Meeting of the
American Physical Society Division of Particles and Fields), 10--14
November 1992, ed. by C. H. Albright \ite~(World Scientific, Singapore,
1993)}
\def \dpf94{DPF 94 Meeting, Albuquerque, NM, Aug.~2--6, 1994}
\def \efi{Enrico Fermi Institute Report No. EFI}
\def \el#1#2#3{Europhys.~Lett.~{\bf#1}, #2 (#3)}
\def \epjc#1#2#3{Eur.~Phys.~J.~C {\bf#1}, #2 (#3)}
\def \f79{{\it Proceedings of the 1979 International Symposium on Lepton
and Photon Interactions at High Energies,} Fermilab, August 23-29, 1979,
ed.~by T. B. W. Kirk and H. D. I. Abarbanel (Fermi National Accelerator
Laboratory, Batavia, IL, 1979}
\def \hb87{{\it Proceeding of the 1987 International Symposium on Lepton
and Photon Interactions at High Energies,} Hamburg, 1987, ed.~by W. Bartel
and R. R\"uckl (Nucl. Phys. B, Proc. Suppl., vol. 3) (North-Holland,
Amsterdam, 1988)}
\def \ib{{\it ibid.}~}
\def \ibj#1#2#3{{\it ibid.}~{\bf#1}, #2 (#3)}
\def \ichep72{{\it Proceedings of the XVI International Conference on High
Energy Physics}, Chicago and Batavia, Illinois, Sept. 6--13, 1972,
edited by J. D. Jackson, A. Roberts, and R. Donaldson (Fermilab, Batavia,
IL, 1972)}
\def \ijmpa#1#2#3{Int.~J.~Mod.~Phys.~A {\bf#1}, #2 (#3)}
\def \ite{{\it et al.}}
\def \jmp#1#2#3{J.~Math.~Phys.~{\bf#1}, #2 (#3)}
\def \jpg#1#2#3{J.~Phys.~G {\bf#1}, #2 (#3)}
\def \lkl87{{\it Selected Topics in Electroweak Interactions} (Proceedings
of the Second Lake Louise Institute on New Frontiers in Particle Physics,
15--21 February, 1987), edited by J. M. Cameron \ite~(World Scientific,
Singapore, 1987)}
\def \KEK#1{{\it Flavor Physics} (Proceedings of the Fourth International
Conference on Flavor Physics, KEK, Tsukuba, Japan, 29--31 October 1996),
edited by Y. Kuno and M. M. Nojiri, Nucl.~Phys.~B Proc.~Suppl.~{\bf 59},
#1 (1997)}
\def \ky85{{\it Proceedings of the International Symposium on Lepton and
Photon Interactions at High Energy,} Kyoto, Aug.~19-24, 1985, edited by M.
Konuma and K. Takahashi (Kyoto Univ., Kyoto, 1985)}
\def \mpla#1#2#3{Mod.~Phys.~Lett.~A {\bf#1}, #2 (#3)}
\def \nc#1#2#3{Nuovo Cim.~{\bf#1}, #2 (#3)}
\def \nima#1#2#3{Nucl.~Instr.~Meth.~A {\bf#1}, #2 (#3)}
\def \np#1#2#3{Nucl.~Phys.~{\bf#1}, #2 (#3)}
\def \npbps#1#2#3{Nucl.~Phys.~B (Proc.~Suppl.) {\bf#1}, #2 (#3)}
\def \pisma#1#2#3#4{Pis'ma Zh.~Eksp.~Teor.~Fiz.~{\bf#1}, #2 (#3) [JETP
Lett. {\bf#1}, #4 (#3)]}
\def \pl#1#2#3{Phys.~Lett.~{\bf#1}, #2 (#3)}
\def \plb#1#2#3{Phys.~Lett.~B {\bf#1}, #2 (#3)}
\def \pr#1#2#3{Phys.~Rev.~{\bf#1}, #2 (#3)}
\def \pra#1#2#3{Phys.~Rev.~A {\bf#1}, #2 (#3)}
\def \prd#1#2#3{Phys.~Rev.~D {\bf#1}, #2 (#3)}
\def \prl#1#2#3{Phys.~Rev.~Lett.~{\bf#1}, #2 (#3)}
\def \prp#1#2#3{Phys.~Rep.~{\bf#1}, #2 (#3)}
\def \ptp#1#2#3{Prog.~Theor.~Phys.~{\bf#1}, #2 (#3)}
\def \rmp#1#2#3{Rev.~Mod.~Phys.~{\bf#1}, #2 (#3)}
\def \rp#1{~~~~~\ldots\ldots{\rm rp~}{#1}~~~~~}
\def \si90{25th International Conference on High Energy Physics, Singapore,
Aug. 2-8, 1990}
\def \slc87{{\it Proceedings of the Salt Lake City Meeting} (Division of
Particles and Fields, American Physical Society, Salt Lake City, Utah,
1987), ed.~by C. DeTar and J. S. Ball (World Scientific, Singapore, 1987)}
\def \slac89{{\it Proceedings of the XIVth International Symposium on
Lepton and Photon Interactions,} Stanford, California, 1989, edited by M.
Riordan (World Scientific, Singapore, 1990)}
\def \smass82{{\it Proceedings of the 1982 DPF Summer Study on Elementary
Particle Physics and Future Facilities}, Snowmass, Colorado, edited by R.
Donaldson, R. Gustafson, and F. Paige (World Scientific, Singapore, 1982)}
\def \smass90{{\it Research Directions for the Decade} (Proceedings of the
1990 Summer Study on High Energy Physics, June 25 -- July 13, Snowmass,
Colorado), edited by E. L. Berger (World Scientific, Singapore, 1992)}
\def \stone{{\it B Decays}, edited by S. Stone (World Scientific,
Singapore, 1994)}
\def \tasi90{{\it Testing the Standard Model} (Proceedings of the 1990
Theoretical Advanced Study Institute in Elementary Particle Physics,
Boulder, Colorado, 3--27 June, 1990), edited by M. Cveti\v{c} and P.
Langacker (World Scientific, Singapore, 1991)}
\def \vanc{29th International Conference on High Energy Physics, Vancouver,
23--31 July 1998}
\def \yaf#1#2#3#4{Yad.~Fiz.~{\bf#1}, #2 (#3) [Sov.~J.~Nucl.~Phys.~{\bf #1},
#4 (#3)]}
\def \zhetf#1#2#3#4#5#6{Zh.~Eksp.~Teor.~Fiz.~{\bf #1}, #2 (#3) [Sov.~Phys.
-- JETP {\bf #4}, #5 (#6)]}
\def \zpc#1#2#3{Zeit.~Phys.~C {\bf#1}, #2 (#3)}

\end{document}